\documentstyle[11pt]{article}

\begin{document}

\baselineskip = 14pt
\vspace*{-0.8cm}
\begin{flushright}
DPNU-01-25
\end{flushright}
\vspace{-0.8cm}
\begin{center}
\Large\bf
Chiral Pertuabtion in the Hidden Local Symmetry and\\
Fate of Vector Dominance
\footnote{%
Talk given at YITP workshop
``Aspects of Particle Physics in New Century''
(July 10 -- 13, 2001, Yukawa Institute, Kyoto, Japan).
This talk is based on the work done in collaboration with
Prof. Yamawaki~\cite{HY:VD}.}
\end{center}
\vspace{-0.5cm}
\begin{flushright}
Masayasu Harada (Nagoya Univ.)
\end{flushright}
\vspace{-0.8cm}
\begin{abstract}
In this talk I summarize our recent
work on the vector dominance in QCD 
by using the the hidden local symmetry as an effective
field theory of QCD.
\end{abstract}

Since Sakurai advocated Vector Dominance (VD)~\cite{Sakurai},
VD has been a widely accepted notion in describing vector meson
phenomena in hadron physics. 
In Ref.~\cite{HY:VD} 
we revealed the full phase structure of the 
hidden local symmetry (HLS)
through the one-loop
renormalization group equation (RGE) 
including quadratic divergences.
We then showed that VD is not a sacred discipline of
the effective field theory but rather an accidental phenomenon
peculiar to three-flavored QCD.

In this talk I summarize the main points of the work 
done in Ref.~\cite{HY:VD} 

Let me start with the successful predictions of the 
HLS~\cite{BKUYY:BKY}.
Lagrangian of the HLS with least derivative terms
includes three
parameters $F_\pi$, $a$ and $g$.
By making a dynamical assumption of a parameter choice $a=2$, 
the HLS predicts the following outstanding phenomenological
facts~\cite{BKUYY:BKY}:
(1) $g_{\rho\pi\pi}=g$ (universality of the
$\rho$-coupling~\cite{Sakurai};
(2) $m_\rho^2 = 2 g_{\rho\pi\pi}^2 F_\pi^2$
(KSRF II)~\cite{KSRF};
(3) $g_{\gamma\pi\pi}=0$ (VD of the
electromagnetic form factor of the $\pi$)~\cite{Sakurai}.
Thus, {\it even though the vector mesons are gauge bosons}
of the HLS, {\it VD is not
automatic} consequence but rather dynamical one of a parameter choice
of $a=2$.

Due to quantum corrections
the parameters change their values by the energy scale,
which are determined by the RGE's.
Accordingly,
values of the parameters $F_\pi$, $a$ and $g$
cannot be freely chosen,
although they are independent at tree level.
Here I stress that {\it thanks to the gauge symmetry}
in the HLS model it is possible to perform
a {\it systematic loop expansion} including the vector mesons in
addition to the pseudoscalar
mesons~\cite{Georgi,HY,Tanabashi,HY:matching,HY:PR}
in a way to extend the chiral perturbation theory~\cite{ChPT}.
Furthermore,
as shown in Ref.~\cite{HY:letter,HY:matching}
it is important to include {\it quadratic divergences}
in calculating the quantum corrections.
Due to quadratic divergences in
the HLS dynamics, it follows that
{\it even if the bare theory defined by the cutoff $\Lambda$
is written as if it were in the broken
phase characterized by $F_\pi^2(\Lambda) > 0$},
{\it the quantum theory can be in the symmetric
phase characterized by $F_\pi^2(0)=0$}~\cite{HY:letter}.

In Ref.~\cite{HY:VD}
we first studied the RG flows of the parameters
and the phase structure of the HLS to classify the parameter space.
Here I skip the explanation of those study on the phase structure of
the HLS
(for details, please see Refs.~\cite{HY:VD,HY:PR}),
and just present the fixed points of the RGE's.
For analysing the RGE's
it is convenient to introduce the 
following quantities:
$X(\mu) \equiv C \mu^2/F_\pi^2(\mu)$ and
$G(\mu) \equiv C g^2(\mu)$,
where $C = N_f/\left[2(4\pi)^2\right]$.
In the RGE's for $X$, $a$ and $G$
there exist
{\it three fixed points} and {\it one fixed line
in the physical region} and one fixed point in the
unphysical region (i.e., $a<0$ and $X<0$).
Those in the physical region (labeled by $i=1,\ldots,4$)
are given by 
\begin{eqnarray}
&& \left(X^\ast_i,\, a^\ast_i,\, G^\ast_i\right)
  = \left( 0,\, \mbox{any},\, 0 \right) \ , 
  \quad
  \left( 1,\, 1,\, 0 \right) \ ,
  \quad
  \left( \frac{3}{5},\, \frac{1}{3},\, 0 \right) \ ,
\nonumber\\
&& \quad
  \left( 
    \frac{2(2+45\sqrt{87})}{4097},\, 
    \sqrt{87},\, 
    \frac{2(11919-176\sqrt{87})}{1069317}
  \right)
\ .
\label{fixed points}
\end{eqnarray}

Let me now discuss the VD which is characterized by $a(0) = 2$,
where $a(0)$ is given by
\begin{equation}
a(0)
=
a(m_\rho) / \left[ 1 + a(m_\rho) X(m_\rho) - 2X(m_\rho) \right]
\ .
\label{def:a0}
\end{equation}
This implies that the VD ($a(0) =2$) is only realized
for $( X(m_\rho), a(m_\rho) ) = 
( 1/2, \mbox{any} )$ or
$( \mbox{any}, 2 )$.

In $N_f=3$ QCD, the parameters at $m_\rho$ scale,
$\left( X(m_\rho), a(m_\rho), G(m_\rho)\right)
\simeq \left( 0.46,  1.22, 0.38 \right)$,
happen to be near such a VD point.
However, the RG flow actually belongs to the fixed point
$\left(X^\ast_4,a^\ast_4,G^\ast_4\right)$ which is far away from the
VD value.
Thus, {\it the VD in $N_f=3$ QCD is accidentally realized by 
$X(m_\rho)\sim1/2$ which is very unstable against the RG flow}
(see Fig.~3 of Ref.~\cite{HY:VD}).
For $G=0$ the VD holds only if the parameters are
(accidentally)
chosen to be on the RG flow entering
$\left( X, a, G \right) = \left( 0, 2, 0 \right)$
which is an end point of the line
$\left( X(m_\rho), a(m_\rho) \right)=(\mbox{any},2)$.
For $a=1$
on the other hand,
the VD point
$\left( X, a, G \right) = \left( 1/2, 1, 1/2 \right)$
lies on the line
$\left( X(m_\rho), a(m_\rho) \right)=(1/2,\mbox{any})$.

The phase diagrams shown in Ref.~\cite{HY:VD}
show that neither 
$X(m_\rho) = 1/2$ nor $a(m_\rho) = 2$ is a special point in the
parameter space of the HLS.
Thus, it was concluded that
the {\it VD can be satisfied only accidentally}.
Then,
when the parameter of QCD is changed,
the VD is generally violated.
In particular, neither 
$X(m_\rho) = 1/2$ nor $a(m_\rho) = 2$ is satisfied on the phase
boundary surface where the
chiral restoration takes place in HLS model.
Therefore,
{\it VD is realized nowhere on the chiral restoration surface !}

Moreover, 
{\it when the HLS is matched with QCD},
only the point
$( X^\ast_2, a^\ast_2, G^\ast_2 ) = ( 1, 1, 0 )$, 
the {\it VM point},
on the phase boundary is selected,
since 
the axialvector and vector current correlators in HLS
can be matched with those in QCD only at that point~\cite{HY:VM}.
{\it
Therefore, QCD predicts $a(0)=1$,
i.e., large violation of the VD at chiral restoration}.
Actually, for the chiral restoration
{\it in the large $N_f$ QCD}~\cite{lattice,ATW}
{\it the VM can in fact
takes place}~\cite{HY:VM}, {\it and thus the VD is badly violated}.

\vspace{-0.2cm}

\paragraph{Acknowledgment}
\ \par

I would like to thank Professor Koichi Yamawaki for collaboration in
Ref.~\cite{HY:VD} on which this talk is based.
I would be grateful to the organizers for giving me 
an opportunity to present this talk.
This work is supported in part by Grant-in-Aid for Scientific Research
(A)\#12740144.

\end{document}